\documentclass[amsmath,amssymb,aps,prd,onecolumn,groupedaddress,nofootinbib]{revtex4-2}

\usepackage{bm}
\usepackage{braket}
\usepackage{graphicx}
\usepackage{mathtools}
\usepackage{dcolumn}
\usepackage{color}
\usepackage[unicode=true,colorlinks=true,linkcolor=magenta, urlcolor=blue, citecolor = blue,breaklinks]{hyperref}

\begin{document}
	
	\title{Wave packet description of Majorana neutrino oscillations in a magnetic field}
	
	\author{Artem Popov}
	\email{ar.popov@physics.msu.ru}
	\affiliation{Department of Theoretical Physics, \ Moscow State University, Moscow 119991, Russia}
	\author{Alexander Studenikin}
	\affiliation{Department of Particle Physics and Extreme States of Matter, Department of Theoretical Physics, Moscow State University, 119991 Moscow, Russia}
	\author{Alexander Tcvirov}
	\affiliation{Department of Theoretical Physics, \ Moscow State University, Moscow 119991, Russia}
\begin{abstract}
Majorana neutrino oscillations in a magnetic field are considered using the wave packets formalism. The modified Dirac equation for Majorana neutrinos with non-zero transition magnetic moments propagating in a magnetic field is solved analytically in the two flavour case. The expressions for the oscillations probabilities are derived accounting for the decoherence effect emerging at distances exceeding the coherence length. It is shown that for Majorana neutrinos propagating in a magnetic field the coherence length coincides with the coherence length for neutrino oscillations in vacuum when the vacuum frequency is much greater than the magnetic frequency ($\omega_{vac} \gg \omega_B$), while it is proportional to the cube of the average neutrino momentum if ($\omega_{vac} \ll \omega_B$). We show that the decoherence effect may appear during neutrino propagation in a magnetic field of supernova.
\end{abstract}

\maketitle


\section{Introduction}\label{intro}
It is well known that massive neutrinos possess non-zero electromagnetic form factors \cite{Giunti:2014ixa,Giunti:2025}, particularly anomalous magnetic moments \cite{Fujikawa:1980yx,Shrock:1982sc}. Presently, the most stringent terrestrial upper bound on neutrino effective magnetic moment is obtained by GEMMA \cite{Beda:2012zz} experiment and is on the level of $2.9 \times 10^{-11} \mu_B$. XENONnT provides the upper bound on the level of $6.4\times10^{-12}\mu_B$ inferred from the solar neutrinos data \cite{XENONCollaboration:2022kmb}. There are also a variety of astrophysical limits of order $10^{-12} \mu_B$ \cite{PDG:2022}. Although neutrinos magnetic moments are extremely small, they may be important to describe neutrino propagation in astrophysical environments with extremely large magnetic fields.

Majorana theory of fermions presents an interesting possibility to describe neutral spin-1/2 particles, in particular neutrinos. Majorana neutrinos naturally appear in some models of physics beyond the Standard Model. As truly neutral particles, Majorana neutrinos can only possess off-diagonal, or transition magnetic moments \cite{Giunti:2014ixa,Giunti:2025}.

When neutrinos traverse sufficiently large distance, the decoherence between the massive neutrinos states may appear as a result of their wave packets propagating with different group velocities. This effect is described within the framework of the wave packet approach to neutrino oscillations. Previously, the wave packet formalism was applied to the problem of vacuum neutrino oscillations \cite{Nussinov:1976uw, Kayser:1981ye, Kiers:1995zj, Egorov:2019vqv, Naumov:2020yyv, Naumov:2013uia}, neutrino oscillations in matter \cite{Peltoniemi:2000nw,Kersten:2015kio} and neutrino collective oscillations during supenova explosions \cite{Akhmedov:2017mcc}. Oscillations of Dirac neutrinos in a magnetic field were considered using the wave packets formalism in \cite{bib2}. In the present paper the wave packet approach is extended to describe Majorana neutrinos propagation in a strong magnetic field.

\section{Majorana neutrinos interaction with a magnetic field}
Majorana neutrinos magnetic moments interactions with a magnetic field is described by the effective Lagrangian
\begin{equation}\label{mag_field_int}
	\mathcal{L}_{B} = -\sum_{ik}\mu_{ik}\overline{(\hat{\nu}_i^L)^c} \bm{\Sigma}\bm{B} \hat{\nu}_k^L + \text{h.c.} = -\sum_{\alpha\beta}\mu_{\alpha \beta}^{(f)} \overline{(\nu_{\alpha}^L)^c} \bm{\Sigma}\bm{B} \hat{\nu}_{\beta}^L  + \text{h.c.},
\end{equation}
where $\mu$ is the hermitian and antisymmetric matrix of the transition magnetic moments in the massive neutrino basis. Here $i,j=1,2,3$; $\alpha,\beta = e,\mu,\tau$ and $\bm{\Sigma}= \begin{pmatrix}
	\bm{\sigma} & 0\\
	0 & \bm{\sigma}
\end{pmatrix}$, where $\bm{\sigma}$ is the vector of Pauli matrices. The flavour neutrino fields operators are given by
\begin{equation}
	\hat{\nu}_{\alpha} = \hat{\nu}_{\alpha}^L + (\hat{\nu}_{\alpha}^L)^c = \sum_{i} U_{\alpha i} \hat{\nu}_i^L + \sum_{i} U_{\alpha i}^* (\hat{\nu}_i^L)^c,
\end{equation}
where $\alpha = e,\mu,\tau$; $i=1,2,3$ and $\hat{\nu}_i^L$ are the chiral fields of massive neutrinos, and $U$ is the neutrino mixing matrix. Majorana neutrinos magnetic moments in the flavour basis $\mu^{(f)}$ are connected to $\mu$ as follows
\begin{equation}\label{mm_flavour}
	\mu^{(f)}=U\mu U^T.
\end{equation}

We suppose that massive neutrinos are described by the wave packets
\begin{equation}
	\ket{i,p_0,h,\sigma_p} = \int \frac{dk}{2\pi} f(k,p_0,\sigma_p)\ket{i,k,h},
\end{equation}
where $\ket{i,k,h} = a^\dag_h(k)\ket{0} \otimes\ket{i}$ is the plane wave state and $a^\dag_h(k)$ is the creation operator of a massive Majorana neutrino with helicity $h$ and momentum $k$. Here the normalization of the states is given by $\braket{j,q,h'|i,p,h} = 2\pi \delta_{ij}\delta_{hh'}\delta(p-q)$ and $\braket{i|j} = \delta_{ij}$. The function $f(k,p_0,\sigma_p)$ corresponds to the wave packet with the average momentum $p_0$ and the momentum spread $\sigma_p$, and satisfies the normalization condition
\begin{equation}
	\int \frac{dk}{2\pi} |f(k,p_0,\sigma_p)|^2 = 1.
\end{equation}
For simplicity, in this paper we consider one dimensional wave packets.

One-particle wave functions of the massive neutrinos states with helicity $h$ are constructed as follows
\begin{equation}\label{one_part_wf}
	\nu_i^h(x) = \bra{i}\otimes\braket{0 | \hat{\nu}_i(x) | i,p_0,h,\sigma_p} = \int \frac{dk}{2\pi} f(k,p_0,\sigma_p) \braket{0|\hat{\nu}_i(x)|k,h},
\end{equation}
and are simply the convolutions of the plane wave one-particle wave functions $\braket{0|\hat{\nu}_i(x)|k,h}$ with the wave packet shape function $f(k,p_0,\sigma_p)$.

Note that only the neutrino field operators $\hat{\nu}_i$ satisfy the Majorana condition $\hat{\nu}_i^c = \hat{\nu}_i$, but not the one-particle wave functions $\nu_i^h(x)$. In the Majorana theory, left-handed states $\nu_i^-$ are usually associated with neutrinos, while right-handed states $\nu_i^+$ are associated with antineutrinos. This is due to the fact that right-handed Majorana neutrinos behave similarly to Dirac antineutrinos in the processes involving neutrino creation and annihilation, as well as in neutrino oscillations.

Using (\ref{mag_field_int}) and (\ref{one_part_wf}), one can derive the system of modified Dirac equations that describes Majorana neutrinos propagating in a magnetic field
\begin{equation} \label{dirac_equation}
	(i\gamma^{\mu} \partial_{\mu} - m_i) \nu_i^h (x) + \sum_{j} \mu_{ij} \bm{\Sigma}\bm{B} \nu_j^h (x) = 0.
\end{equation}
In the next section we analytically solve Eq. (\ref{dirac_equation}) in the two flavour case and calculate the Majorana neutrino oscillations probabilities accounting for decoherence due to the neutrinos wave packets separation.

\section{Majorana neutrino oscillations in a magnetic field}
We solve the Dirac equations system (\ref{dirac_equation}) assuming that neutrino propagates along $z$-axis and is initially in a state with definite helicity $h$. In this case the initial wave functions of the massive neutrino states are given by
\begin{equation}
	\nu_i^h(z,0) = \int \frac{dk}{2\pi} f(k,p_0,\sigma_p) e^{ikz} u^{h}_i(k),
\end{equation}
where $u_i^h(k)$ is the stationary solution for the free Dirac particle with mass $m_i$, helicity $h = \pm1$ and positive energy. Due to the interaction with a magnetic field, the wave functions $\nu_i^h(z,t)$ for $t>0$ is not a state with definite helicity, but rather a superposition of helicity states.

We assume that the massive neutrino states wave packets are described by the gaussian function
\begin{equation}
	f(k,p_0,\sigma_p) = \frac{(2\pi)^{1/4}}{\sqrt{\sigma_p}}\exp\left(-\frac{(k-p_0)^2}{4\sigma_p^2}\right),
\end{equation}
where $p_0$ is the average neutrino momentum and $\sigma_p$ is the momentum spread.

Eq. (\ref{dirac_equation}) can be rewritten in the Hamiltonian form as follows
\begin{equation}\label{equation_hamiltonian}
	i\frac{\partial}{\partial t} \nu^h(z,t) =  H \nu^h(z,t),
\end{equation}
where $\nu^h = (\nu_1^h, \nu_{2}^h)^T$ is a columns of the neutrino states and the Hamiltonian is defined as a block matrix $H_{ik}$
\begin{equation}\label{hamiltonian}
	H_{ij}=\delta_{ij}\gamma_{0} \gamma_3 k + m_{ij}\delta_{ij}\gamma_{0}-\mu_{ij}\gamma_{0}\bm{\Sigma}\bm{B}.
\end{equation}
For Majorana neutrinos the transition magnetic moments are given by $\mu_{12} = i\mu$ and $\mu_{21} = -i \mu$, where $\mu$ is a real number.

The positive eigenvalues of the Hamiltonian (\ref{hamiltonian}) have the following form
\begin{widetext}
\begin{equation}\label{eigenvals}
	E_s(k) = \sqrt{ p^2 + \frac{m^2_1 + m^2_2}{2} + 2sp\sqrt{\omega_B^2 + \omega_{vac}^2 + \frac{\omega_B^2 (m_1+m_2)^2}{2k^2}} },
\end{equation}
\end{widetext}
where $s=\pm 1$, $\omega_B$ and $\omega_{vac}$ are the vacuum and magnetic oscillations frequencies given by
\begin{equation}
	\omega_{vac}(k) = \frac{\Delta m^2}{4 k}, \;\;\; \omega_B = \mu B_\perp.
\end{equation}

In the limit of the vanishing magnetic field $B \to 0$ eigenvalues (\ref{eigenvals}) reproduce the customary expressions for the ultra-relativistic neutrinos dispersion relation in vacuum
\begin{equation}
	E_+ \approx k + \frac{m_2^2}{2k}, \;\;\; E_- \approx k + \frac{m_1^2}{2k}.
\end{equation}

To simplify further calculations, we make the following assumptions. First, we consider only ultra-relativistic neutrinos, i.e. $p_0 \gg m_i$. Secondly, we assume that $m_i\gg\mu B_\perp$, which, given the current bounds on neutrino mass and magnetic moments, is justified even for extremely strong astrophysical magnetic fields. Under these assumptions, one can decompose the dispersion relation (\ref{eigenvals}) as follows
\begin{widetext}
\begin{eqnarray}\label{eig_approx}
	E_s(k) \approx k \Bigg[ 1 + \frac{m_1^2 + m_2^2}{4k^2} +\frac{s}{k} \sqrt{\omega_{vac}^2 + \omega_B^2 + \frac{\omega_B^2(m_1^2+m_2^2)}{2k^2}} - \frac{1}{8} \Big( \frac{m_1^2 + m_2^2}{2k^2} \Big)^2 \\ \nonumber
	- \frac{1}{2k^2}\Big(\omega_{vac}^2 + \omega_B^2 + \frac{\omega_B^2(m_1^2+m_2^2)}{2k^2}\Big) + \frac{s}{4k^3} (m_1^2 + m_2^2)\sqrt{\omega_{vac}^2 + \omega_B^2 + \frac{\omega_B^2(m_1^2+m_2^2)}{2k^2}} \Bigg].
\end{eqnarray}
\end{widetext}

Using (\ref{eig_approx}), we calculate the difference between group velocities $v_+ = \dfrac{\partial E_+}{\partial k}$ and $v_- = \dfrac{\partial E_-}{\partial k}$ of the Majorana neutrino stationary states in a magnetic field
\begin{widetext}
\begin{equation}\label{group_vel_diff}
	\Delta v = v_+ - v_- \approx \frac{1}{k^3 \sqrt{\omega_{vac}^2 + \omega_B^2}} \Bigg( \frac{(\Delta m^2)^2}{8} + \omega_B^2 (m_1 + m_2)^2 \Bigg) + \frac{(m_1+m_2)^2}{4 p^3} \sqrt{\omega_{vac}^2 + \omega_B^2}.
\end{equation}
\end{widetext}
The group velocities difference $\Delta v$ characterizes the wave packets separation rate and the resulting neutrino decoherence.

The expression for the group velocity difference (\ref{group_vel_diff}) is significantly simplified in two asymptotic regimes $\omega_{B} \ll \omega_{vac}$ and $\omega_{B} \gg \omega_{vac}$
\begin{eqnarray}\label{group_vel_diff_asympt}
	\Delta v &\approx& \frac{\Delta m^2}{2k^2} \;\;\; \text{if } \omega_{vac} \gg \omega_B, \\ \nonumber
	\Delta v &\approx& \frac{(\Delta m^2)^2/\omega_B + 2 \omega_B (m_1+m_2)^2}{k^3}  \;\;\; \text{if } \omega_{vac} \ll \omega_B.
\end{eqnarray}
In the case $\omega_{B} \ll \omega_{vac}$ it coincides with the group velocities difference for neutrinos propagating in vacuum.

Using the approach developed in \cite{bib1}, we calculate the probabilities of transitions between neutrinos $\nu_\alpha \to \nu_\beta$ (that are left-handed neutrinos $\nu^-_\alpha$ and $\nu^-_\beta$ in the Majorana case) and between neutrinos and antineutrinos $\nu_\alpha \to \bar{\nu}_\beta$ (transitions between left- and right-handed neutrino states $\nu^-_\alpha$, and $\nu^+_\beta$ in the Majorana case) of different flavours:
\begin{widetext}
	\begin{eqnarray}\label{p1}
		P(\nu_e \rightarrow \nu_\mu;L) &=& \frac{\omega_{vac}^2}{2(\omega_{vac}^2 + \omega_B^2)} \sin^22\theta \left[1 + \cos\left(\frac{2\pi L}{L_{osc}}\right) \exp\left(-\frac{L^2}{L_{coh}^2} \right) \right], \\
		\label{p2}
		P(\nu_e \rightarrow \bar{\nu}_\mu;L) &=& \frac{\omega_B^2}{2( \omega_{vac}^2 + \omega_B^2 )}\left[1 + \cos\left(\frac{2\pi L}{L_{osc}}\right) \exp\left(-\frac{L^2}{L_{coh}^2} \right) \right], \\
		\label{p3}
		P(\nu_e \rightarrow \bar{\nu}_e;L) &=& 0,
	\end{eqnarray}
\end{widetext}
	where
	\begin{equation}
		L_{osc} = \frac{2\pi}{\sqrt{\omega_B^2 + \omega^2_{vac}}}
	\end{equation}
	is the oscillations length, and
	\begin{equation}\label{coh_length_exact}
		L_{coh} = \frac{2\sqrt{2}\sigma_x}{\Delta v(p_0)},
	\end{equation}
	is the coherence length that characterizes the spatial scale on which the damping of the neutrino transitions occurs. Here $\sigma_x = 1/2\sigma_p$ is the wave packet width in the coordinate space.
	Eqs. (\ref{p1}) generalize the expressions obtained in \cite{Kurashvili:2017zab} in the plane wave case and accounts for the damping of neutrino oscillations. In our calculations we assumed that the Majorana CP-violating phase is absent. Possible effects of non-zero Majorana CP-violating phases in neutrino oscillations in a magnetic field were considered in \cite{bib1,Lichkunov:2025rpu}
	
	Using (\ref{group_vel_diff_asympt}), one can obtain the asymptotic expressions for the coherence length
	\begin{eqnarray}\label{coh_length_asympth}
		L_{coh} &\approx& \frac{4\sqrt{2}\sigma_x p_0^2}{\Delta m^2} \;\;\; \text{if } \omega_{vac} \gg \omega_B, \\ \nonumber
		L_{coh} &\approx& \frac{2\sqrt{2} \sigma_x p_0^3}{(\Delta m^2)^2/\omega_B + 2 \omega_B (m_1+m_2)^2}  \;\;\; \text{if } \omega_{vac} \ll \omega_B.
	\end{eqnarray}
	If the vacuum frequency $\omega_{vac} = \Delta m^2/4p_0$ is much greater then the magnetic frequency $\omega_B = \mu B_\perp$, the oscillations probabilities (\ref{p1}) and the coherence length (\ref{coh_length_exact}) coincide with those for neutrino oscillations in vacuum. In this case neutrino-antineutrino transitions induced by a magnetic field are strongly suppressed. In the opposite case $\omega_B \gg \omega_{vac}$, the coherence length is proportional to the cube of the average neutrino momentum $(L_{coh} \sim p_0^3)$. The same dependence holds for the case of Dirac neutrinos oscillations in a magnetic field \cite{bib2}, although the exact expression is different. The neutrino-neutrino transitions are suppressed if $\omega_B \gg \omega_{vac}$. In the intermediate case $\omega_B \sim \omega_{vac}$, the coherence length exhibits a more complicated non-power-law dependence on the average neutrino momentum.
	
	Future neutrino experiments JUNO, Hyper-Kamiokande, DUNE and others are going to be sensitive to neutrinos produced during the supernovae explosions in our or neighbor galaxies. Strong magnetic fields $10^{12}$ Gauss and even more that occur during supernova explosions may influence neutrino oscillations. Below we consider possible decoherence effects in oscillations of neutrinos in supernova magnetic field.
	
	The coherence and oscillations lengths for the case of Majorana neutrino oscillations in a magnetic field of a supernova explosion are shown in Figure 1 as functions of neutrino energy. Since we consider ultrarelativistic neutrinos, the average momentum $p_0$ approximately equals neutrino energy. We assume that the magnitude of the Majorana neutrino transition magnetic moment is $\mu = 10^{-12} \mu_B$, that is lower than the current experimental upper-bound and may be achieved e.g. in supersymmetric models \cite{Zhang:2024ijy}. We compare the case of $B=10^{12}$ Gauss (left) and $B=10^{11}$ Gauss (right). We also assume that $\sigma_x = 10^{-12}$ cm according to \cite{Akhmedov:2017mcc}. For realistic supernova neutrinos with energies $\sim 10$ MeV, the coherence lengths are $\sim 10 - 100$ kilometers for atmospheric mass square difference, which is comparable to the size of the supernova region where strong magnetic fields are observed. Thus, we can expect that the decoherence effects may indeed be important to describe the flavour evolution of supernova neutrinos and lead to the damping of both flavour and neutrino-antineutrino transitions. For a considered choice of supernova magnetic field and neutrino magnetic moment, the regime $\omega_B \gg \omega_{vac}$ is realized, and the coherence lengths grow as $E^3$ with increasing neutrino energy, which is a distinctive signature for Majorana neutrinos propagating in a magnetic field.
	
	A more detailed analysis of decoherence effects in neutrino oscillations in supernovae magnetic fields require to account for neutrino interaction with matter, as well as to adopt a realistic model of magnetic field. This is a topic of our future research.
	
	\begin{figure}[tbp]
		\centering 
		\includegraphics[width=.49\textwidth]{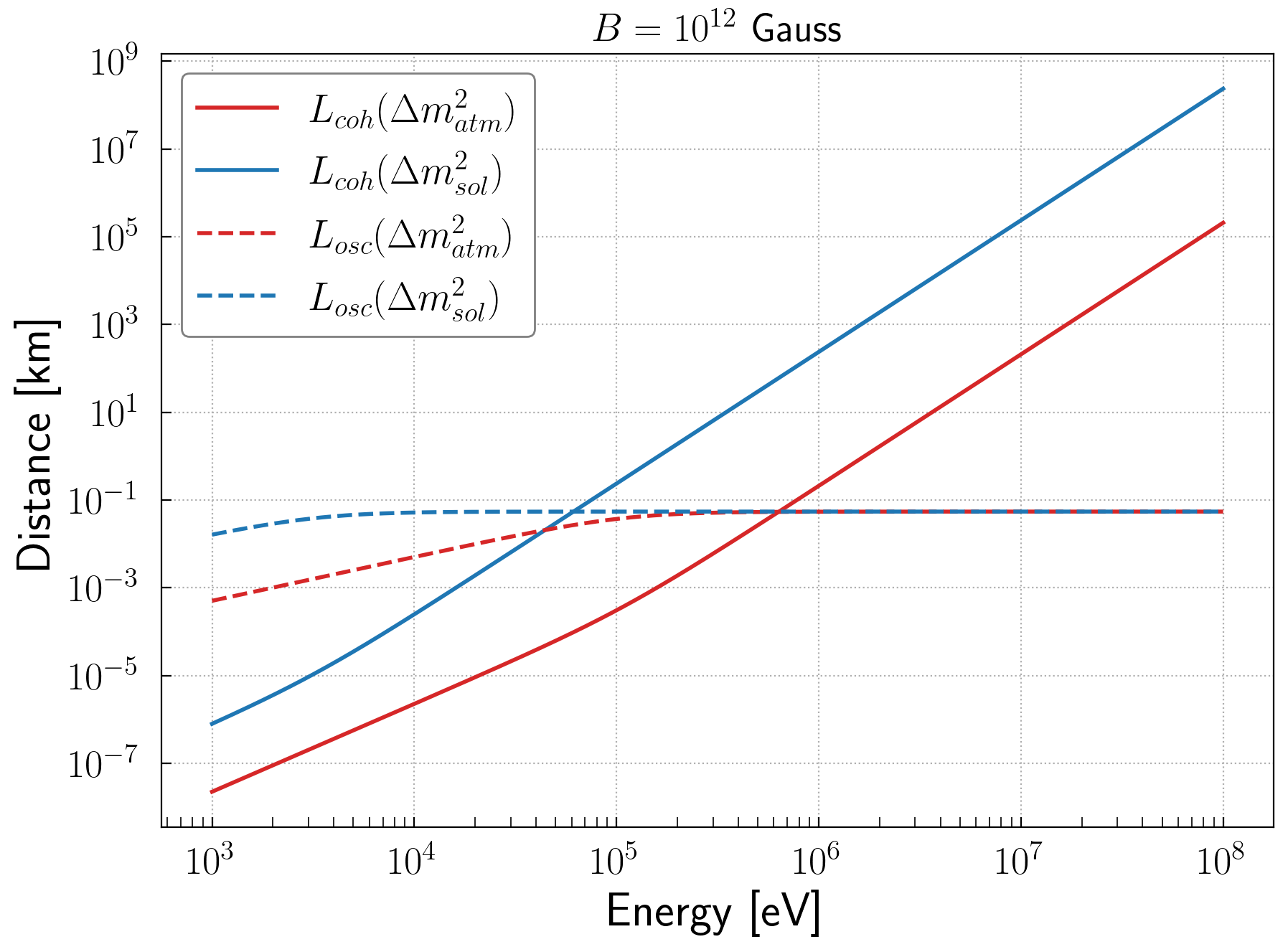}
		\includegraphics[width=.49\textwidth]{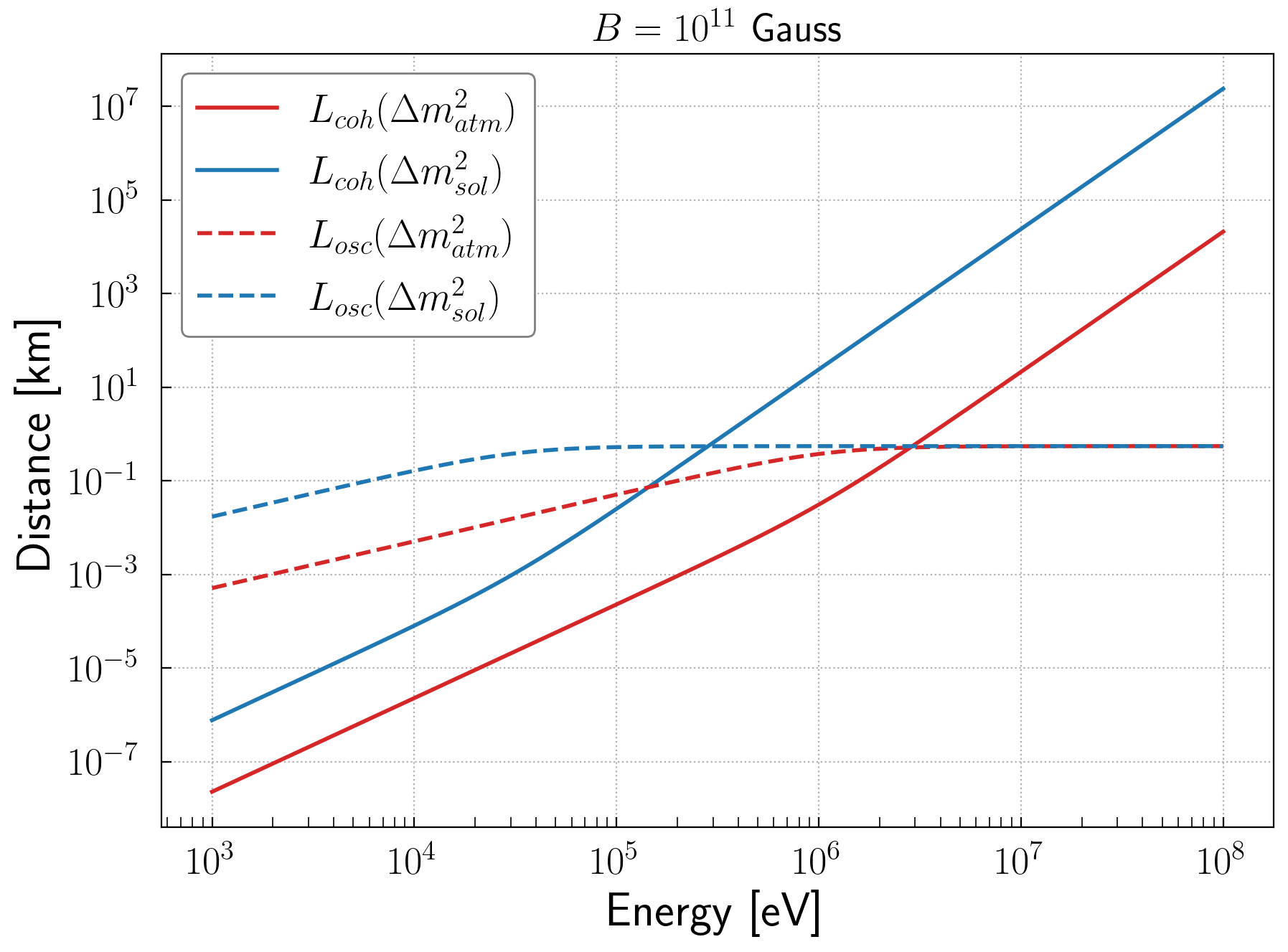}
		\caption{\label{fig:lengths} Coherence and oscillations lengths for Majorana neutrino oscillations in a magnetic field as functions of neutrino energy assuming $\sigma_x = 10^{-12}$ cm. Left: $B= 10^{12}$ Gauss. Right:  $B= 10^{11}$ Gauss.}
	\end{figure}
	
	\section{Conclusion}
	We considered Majorana neutrinos oscillations in strong magnetic fields that appear in astrophysical objects, e.g. during supernova explosions. For the first time we accounted for the decoherence effect due to the neutrino wave packets separation. The analytical expressions for the oscillations probabilities and the coherence length were obtained for the case of two neutrino flavours. It is shown that in the case when $\omega_B \gg \omega_{vac}$, the coherence length is proportional to the cube of the neutrino energy $E^3$ and depends on the magnitude of the magnetic field, while in the limit $\omega_B \ll \omega_{vac}$ the expression for the coherence length coincides with that obtained for the case of neutrino oscillations in vacuum. From our numerical estimations, it follows that the decoherence of Majorana neutrinos oscillations due to wave packets separation may occur in strong magnetic fields of astrophysical objects, such as supernovae.
	
\begin{acknowledgments}
The work is supported by the Russian Science Foundation under grant No.24-12-00084. The work of A.P. has been supported by the National Center for Physics and Mathematics (Project “Study of coherent elastic neutrino-atom and -nucleus scattering and neutrino electromagnetic properties using a high-intensity tritium neutrino source”).
\end{acknowledgments}
	
\section*{CONFLICT OF INTEREST}
The authors declare that they have no conflicts of interest.


\end{document}